\documentclass[12pt,preprint]{aastex}
\usepackage{graphicx}
\shorttitle{Multiple Stellar Populations in GCs}
\shortauthors{Milone et al.}

\newcommand{\hst}{{\it HST\/}}

\begin{document}
\title{The ACS Survey of Galactic Globular Clusters.  III. The Double
Subgiant Branch of NGC 1851\thanks{Based on observations with the
NASA/ESA {\it Hubble Space Telescope}
under the programs GO-10775 and GO-5696,
obtained at the Space Telescope
Science Institute, which is operated by AURA, Inc., under NASA
contract NAS 5-26555.}}

\author{ A.\ P.\ Milone\altaffilmark{1},
L.\ R.\ Bedin\altaffilmark{2},
G.\ Piotto\altaffilmark{1},
J.\ Anderson\altaffilmark{3},
I.\ R.\ King\altaffilmark{4},
A.\ Sarajedini\altaffilmark{5},
A.\ Dotter\altaffilmark{6},
B.\ Chaboyer\altaffilmark{6},
A.\ Mar\'\i n-Franch\altaffilmark{7},
S.\ Majewski\altaffilmark{8},
A.\ Aparicio\altaffilmark{7},
M.\ Hempel\altaffilmark{5},
N.\ E.\ Q.\ Paust\altaffilmark{2},
I.\ N.\ Reid\altaffilmark{2},
A. Rosenberg\altaffilmark{7},
M.\ Siegel\altaffilmark{9}}
\altaffiltext{1} {Dipartimento  di   Astronomia,  Universit\`a  di
  Padova, Vicolo dell'Osservatorio 3, Padova, I-35122, Italy}
%email{antonino.milone-giampaolo.piotto@unipd.it }
\altaffiltext{2} {Space Telescope Science Institute, 3700 San Martin Drive,
Baltimore, MD 21218, USA}
\altaffiltext{3}{Department of Physics and Astronomy, Mail Stop 108,
	     Rice University, 6100 Main Street, Houston, TX 77005, USA}
%              \email{jay@eeyore.rice.edu}
\altaffiltext{4}{Department of Astronomy, University of Washington,
	      Box 351580, Seattle, WA 98195-1580, USA}
%	      \email{king@astro.washington.edu}
\altaffiltext{5}{Department of Astronomy, University of Florida,
	      211 Bryant Space Science Center, Gainesville, FL 32611, USA}
\altaffiltext{6}{Department of Physics and Astronomy, Dartmouth College,
	      6127 Wilder Laboratory, Hanover, NH 03755, USA}
\altaffiltext{7}{Instituto de Astrof\`\i sica de Canarias, E-38200 La
	      Laguna, Canary Islands, Spain}
\altaffiltext{8}{Dept. of Astronomy, University of Virginia,
P.O. Box 400325, Charlottesville, VA 22904-4325}
\altaffiltext{9}{University of Texas, McDonald Observatory, 1
	      University Station, C1402, Austin TX, 78712}
%             \and
%             }
% }
%\date{Received Xxxxx xx, xxxx; accepted Xxxx xx, xxxx}
%__________________________________________________________________
%

\begin{abstract}
Photometry with {\sl HST\/}'s ACS reveals that the subgiant branch (SGB)
of the globular cluster NGC 1851 splits into two well-defined
branches. If the split is due only to an age effect, the two SGBs
would imply two star formation episodes separated by $\sim$ 1 Gyr. We
discuss other anomalies in NGC 1851 which could be interpreted in
terms of a double stellar population.  Finally, we compare the case of
NGC 1851 with the other two globulars known to host multiple
stellar populations, and show that all three clusters differ in
several important respects.
\end{abstract}

%________________________________________________________________
%

%%%
%
\section{Introduction}
%
%%%

For many decades, globular clusters (GC) have been considered the
simplest possible stellar populations, made up of stars located at the
same distance, formed at the same epoch and from the same material.
Although anomalies had been noted from time to time in the abundances
of a number of individual elements
(see discussion in Gratton, Sneden, \& Carretta 2004), 
the description of populations in terms of a
helium abundance and an overall heavy-element abundance seemed firm.
The ``second-parameter problem'' has continued to be a pain for many
decades, and unusual HB morphologies were turning up more often, but the
implications for the origin and evolution of GC stars remained
ambiguous.
Similarly, theoreticians had predicted that self-enrichment 
(Cotrell \& Da Costa 1981, Ventura et al.\ 2001, Ventura \& D'Antona 2005, Maeder \& Meynet 2005)
or mergers might generate multiple populations in clusters, but such
predictions remained controversial 
(see, e.g., Fenner et al.\ 2004, Bekki \& Norris 2006), 
and had little observational basis.
Most important for our discussion here, the idea of one clear sequence
in each part of the HR diagram of each cluster stood firm.

Now, however, the paradigm of GC hosting simple stellar populations
has been seriously challenged by the discovery of multiple
evolutionary sequences 
in $\omega$~Centauri (Bedin et al.\ 2004)
and NGC 2808 (Piotto et al.\ 2007, P07).  In both clusters the
evidence that stars must have formed in distinct bursts is the
presence of multiple main sequences (MSs). However, the two objects
differ in at least two important aspects: The stars of $\omega$ Cen
have a large spread in metal content, whereas in NGC 2808 only oxygen
and sodium are observed to vary much.  Omega Cen shows at least four
distinct subgiant branches (SGBs), implying a range of ages, while the narrow
turnoff in NGC 2808 implies that there is little or no difference in
age among its populations.  One more cluster, M54, shows a complex
color-magnitude diagram (see, e.g., Layden \& Sarajedini 2000),
including a bifurcated SGB (see paper IV of this series,
Siegel et al.\ 2007).  This cluster has been shown,
however, in too many papers to cite here, to be a part of the
Sagittarius dwarf galaxy that is in process of merging into the Milky
Way, and very possibly the actual nucleus of that galaxy.  (Actually,
it is still matter of debate which parts of the color-magnitude
diagram of M54 represent the cluster population and which ones are due to the
Sagittarius stars.)  Even though $\omega$ Cen could very well
represent a similar situation, we feel that M54 is very different from
the clusters that we discuss here, and we will therefore not include
it in the discussions of this paper.

The puzzling observational facts for $\omega$~Cen and NGC 2808
call for a more careful analysis of the MS, turnoff (TO), and subgiant
branch of other GCs.  In this respect, the color-magnitude
diagram (CMD) database from the \hst\ Treasury program GO-10755 (see
Sarajedini et al.\ 2007) provides a unique opportunity to search for
anomalies in the different evolutionary sequences of other Galactic
GCs.
Even though the narrow F606W $-$ F814W color baseline
is far from ideal for identifying multiple main sequences, our
observing strategy was devised to have a very high signal/noise ratio
at the level of the TO, and therefore the CMDs are perfectly suitable
for identifying multiple TO/SGBs, such as those as found in $\omega$~Cen.
(New \hst\ observations specifically devoted to the identification of
multiple MSs are already planned for other massive GCs, in
GO-10922 and GO-11233.)

Indeed, a first look at the entire Treasury database showed at least
one other GC whose CMD clearly indicates the presence of multiple
stellar populations: NGC 1851.  Although NGC 1851 is a massive GC with
a prototypical bimodal horizontal branch (HB), not much attention has
been devoted to the study of its stellar population. The most complete
photometric investigation of this cluster is by Saviane et al.\
(1998), and the most extended spectroscopic analysis was done over 25
years ago, by Hesser et al.\ (1982).  Saviane et al.\ (1998) confirmed
the bimodal nature of the HB of NGC 1851, but did not find any other
anomaly in its CMD. Hesser et al.\ (1982) found that three out of
eight bright red giant branch (RGB) stars have extremely strong CN
bands.

The present paper is based on new photometry of NGC 1851, from
\hst/ACS imaging and from archival WFPC2 images.  Section 2 describes
the data sets, and \S\ 3 presents the color-magnitude diagram from the
ACS data, with a split in the subgiant branch clearly visible.  In \S\
4 we use the WFPC2 data, both for proper-motion elimination of field
stars and to construct a CMD with a broader wavelength baseline which
we use to set a severe upper limit on any spread in the main sequence.
Section 5 discusses the properties of the SGB and of the horizontal
branch, and shows that the two parts of each have the same spatial
distribution.  Section 6 considers the ages of the two populations,
and \S\ 7 is a discussion and summary.

%__________________________________________________________________
%
\section{Observations and Measurements}
The main database used in this paper comes from {\sl HST} ACS/WFC
images in the F606W and F814W bands, taken for GO-10775 (P.I.\ Sarajedini).
In addition, we use {\sl HST} WFPC2 archive data from GO-05696 (P.I.\
Bohlin) to extend the wavelength range to the blue, as well as to
obtain proper motions. The data sets are summarized in Table 1.

%\begin{center}
\begin{table}
%\centering
\label{t1}
\begin{tabular}{cccc}
\hline\hline  DATE & EXPOSURES & FILT & PROGRAM\\
\hline
%-------------------------------------------------------------------------------------
\multicolumn{4}{c}  {WFPC2}\\
\hline
   April 10, 1996  & 4$\times$900s                   & F336W & 5696 \\
\hline
\multicolumn{4}{c}  {ACS/WFC}\\
\hline
   July 1, 2006  & 20s$+$5$\times$350s               & F606W & 10775  \\
   July 1, 2006  & 20s$+$5$\times$350s               & F814W & 10775  \\
%----------------------------------------------------------------------
\hline
\end{tabular}
\caption{Description of the data sets used in this work.}
\end{table}
%\end{center}

The photometric and astrometric measurements were made for both WFPC2
and ACS using the algorithms described by Anderson \& King (2000, 2006).
Each image was reduced independently, with the observations averaged to
produce a single flux for each star in each filter.
We put the WFC photometry into the Vega-mag system following the
procedures in Bedin et al.\ (2005) and adopting the encircled energy
and zero points as given by Sirianni et al.\ (2005).  WFPC2 data were
photometrically calibrated following Holtzman et al.\ (1995).  The
proper motions were derived as in Bedin et al.\ (2006).

Breathing can change the focus of the telescope, which can result in
small spatial variations of the PSF relative to the library PSF, and
consequently in small spatial variations of the photometric zero
point.  In order to deal with any residual PSF variation, we used a
method similar to the one described in Sarajedini et al.\ (2007) to
perform a spatial fit to the color residuals relative to the
main-sequence ridge line, and remove them.  In this case,
however, the corrections were made only to colors, rather than along a
reddening line, because, as noted in Section 3, the reddening of NGC
1851 is so small that what we are correcting is surely differences
between the spatial variations of the PSF in the two filters, rather
than differential reddening.  The spatially dependent correction was
at the level of $\sim$0.01 magnitude, and merely sharpens the
sequences a little.

\section{The ACS/WFC Color-Magnitude Diagram}
%
%%%

%__________________________________________________________________
%
\begin{figure}
\epsscale{.80}
\plotone {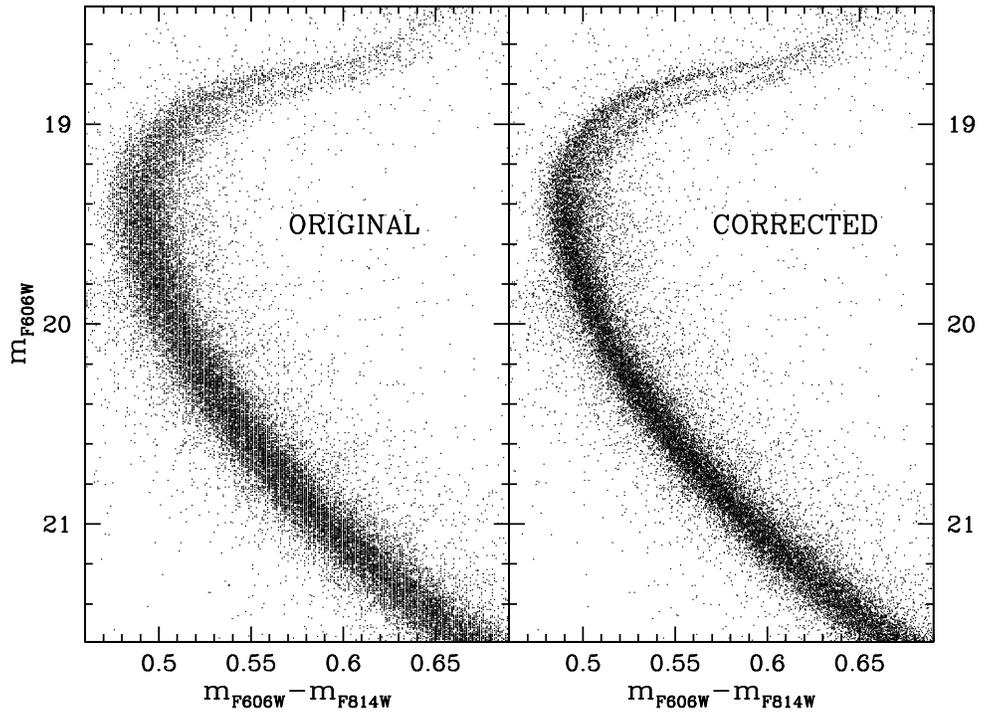}
\caption{{\it Left}: Original CMD. {\it Right}: CMD corrected for spatial
         variation of the photometric zero point of color.
}
\label{f1}
\end{figure}
%__________________________________________________________________

In Figure~\ref{f1} we show the color-magnitude diagram for the
ACS/WFC photometry.  The left panel shows the original CMD.  The main
result is already evident:\ the subgiant branch (SGB) is split into two
distinct branches. The right panel shows the CMD after the correction
for the spatial variation of the color zero point. The SGB split
is still there, and there are clearly two sequences down to the
cluster TO. No separation is evident in the main sequence. It must be
immediately noted that because of the narrowness of MS below the TO,
the SGB broadness can be due neither to any residual spatial variation
of the zero point nor to any differential reddening, which is expected
to be negligible because of the low reddening of NGC 1851
[$E(B-V)=0.02$, Harris 1996].

The two SGBs remain separated by $\sim 0.1$\ mag in the magnitude
interval $18.0<m_{\rm F606W}<19.0$, over the color interval $0.5<m_{\rm
F606W}-m_{\rm F814W}<0.6$.  Hereafter we will refer to the two SGBs
as bSGB and fSGB, where b stands for brighter and f stands for fainter.

Omega Cen and NGC 2808 also show evidence of multiple populations,
but we note that the split in the SGB of NGC 1851 is quite different
from what we found in the former two cases.  In NGC 2808, the main
sequence splits into three distinct sequences a couple of magnitudes below
the turnoff, but the SGB shows no evidence of splitting.  In Omega Cen,
both the SGB and the MS show evidence of splitting.  Here, it is only
the SGB that is split.  We will see that the main sequence of NGC1851
appears consistent with a single population.

%%%
%
\section{The WFPC2 Data}

%%%
%
\subsection{Proper motions for field decontamination}
%
%%%

We used the archival WFPC2 images as a first epoch to determine proper
motions.  WFPC2 has a field of view that is only $\sim$50\% of the ACS
field, and the crowding prevented us from measuring good WFPC2
positions for stars within 500 ACS/WFC pixels of the cluster center.
This somewhat reduces the size of our sample.

\begin{figure}
\epsscale{.80}
   \plotone{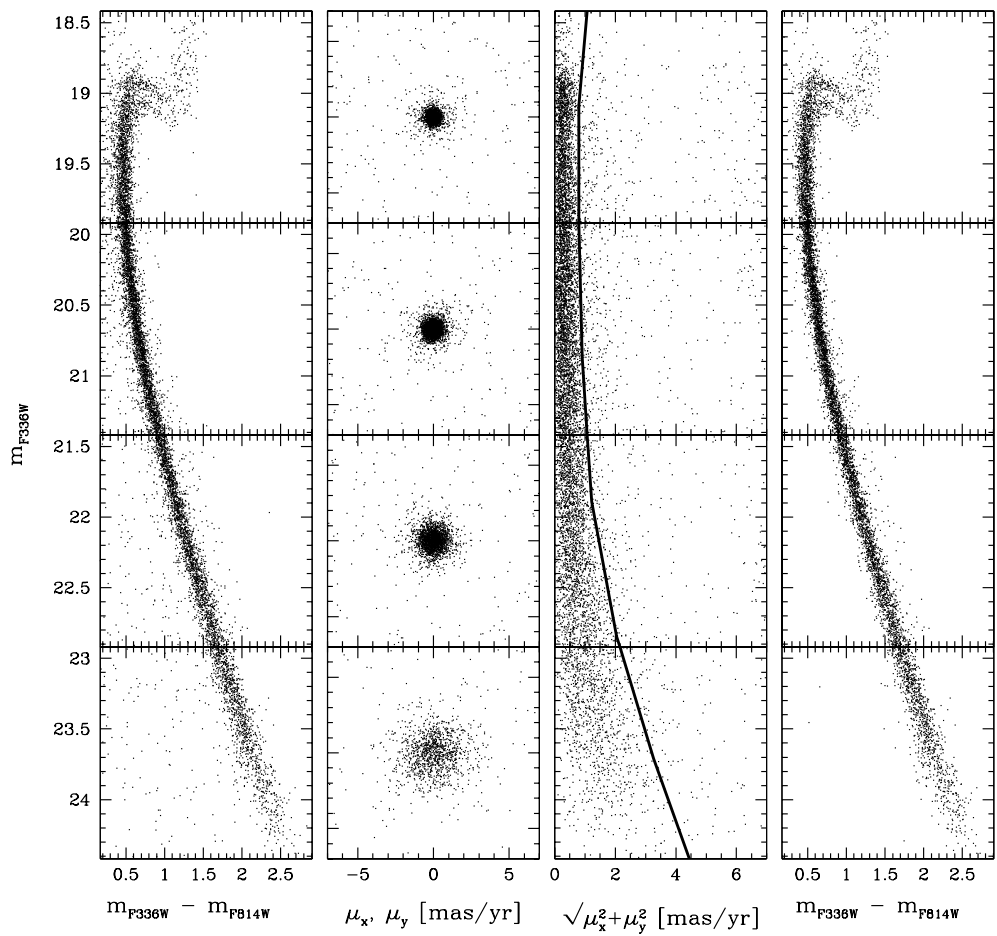}
      \caption{{\it Left}: F336W vs.\ F336W $-$ F814W CMD for all stars.
               {\it Second column}: Proper-motion
               diagram of the stars in the left panel, in intervals
               of 1.5 magnitudes.  {\it Third column}:  The
               total proper motion relative to the cluster mean proper
               motion.   {\it Right}: The
               proper-motion-selected CMD. }
         \label{f2}
   \end{figure}

In the left panel of Fig.~\ref{f2} we show the CMD, using the F336W
magnitudes from the WFPC2 images and the F814W magnitudes from the ACS
images.  The second column of panels shows the proper-motion diagrams
of the stars for four different magnitude intervals.  It should be
noted that we measure proper motions relative to a reference frame
made up of only cluster members, so that the zero point of our motions
is the mean motion of the cluster.  In the third column the magnitudes
of the proper motion vectors are plotted against stellar magnitude.
The errors clearly increase toward fainter magnitudes.  The line was
drawn in order to isolate the stars that have member-like motions, and
was derived as follows.  First, we note that in our study of SGB stars
it is more important to have a pure cluster sample than a complete
one, so we make a conservative choice.  Taking each interval of 1
magnitude in F336W, we begin by estimating the sigma of the member-like
motions.  To do this we find the size of proper motion that includes
70\% of the stars; for a bivariate Gaussian this radius should lie at
$1.552\sigma$, where $\sigma$ refers to each dimension.  We then plot
a point at $2.448\sigma$, which is the size that should include 95\%
of the cluster stars.  We use spline interpolation to draw a line
through these points, and choose all the stars to the left of the
line.  By giving up 5\% of the cluster stars we effectively exclude
nearly all of the field stars.  The rightmost panel shows the CMD of
our chosen cluster stars.

\begin{figure}
\epsscale{.80}
   \plotone{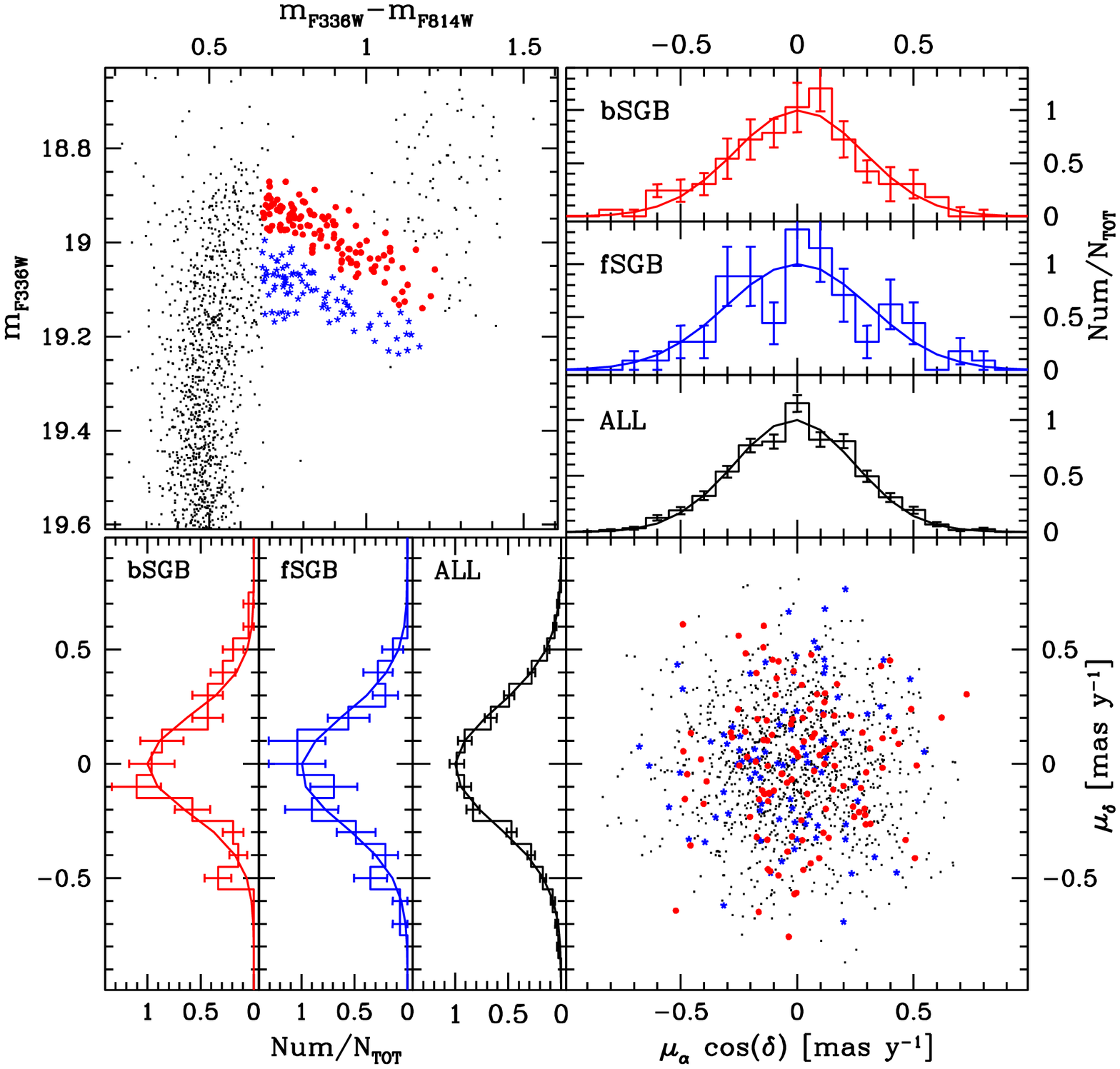}
      \caption{{\it Top left}: The same CMD as in Fig.~\ref{f2}, zoomed
               around the SGB, bSGB stars in red, fSGB stars in blue.
               {\it Bottom right:} proper motions, with same color
               coding.  The other two panels show histograms of the
               individual components of proper motion.}
         \label{f3}
   \end{figure}

In Fig.~\ref{f3} we summarize the results.  The top left panel shows a
zoomed CMD around the SGB for the stars in our proper-motion-selected
sample. Stars marked in red are are bSGB stars (119 objects), and the
ones in blue are fSGB stars (88 objects).  In the bottom right panel is
the vector point diagram for the same stars.  The two SGBs seem to have
the same proper motion distribution.  To show this more clearly, in the
adjacent panels we plot normalized histograms of the separate components
of the proper motions, for the stars of the three groups separately.
Gaussian fits are also shown.

%__________________________________________________________________
%__________________________________________________________________
%

%__________________________________________________________________
%

A side-benefit of proper-motion cleaning is that it also removes from
the CMD stars
that did not have well-measured positions.  These stars tend to have
poor photometry as well, so removing them naturally improves the quality
of the CMD.

%%%
%
\subsection{Wide color baseline}
\label{ss_widecolor}
%
%%%

The second benefit of the WFPC2 data is that it provides a much better
color baseline than the ACS data alone; our study of NGC 2808 (P07)
showed that a wider color baseline yields a more significant splitting
of the main sequence.  The CMD in Figure~\ref{f2} shows F336W vs.\ the
F336W $-$ F814W color.  The SGB split is even more clear in this figure
than in the $V-I$ CMD of Figure~\ref{f1} from the ACS data alone --- but
what is more important is that we now have a better opportunity to study
any possible color spread in the main sequence.  Any limits on the
intrinsic width of the main sequence will translate directly into upper
limits on spreads in the helium and heavy-element abundances, since age
differences have little effect below the turnoff.  (We note, however,
the possibility that simultaneous changes in He and heavy elements could
have effects that offset each other.)

\begin{figure}
 \epsscale{.80}
   \plotone{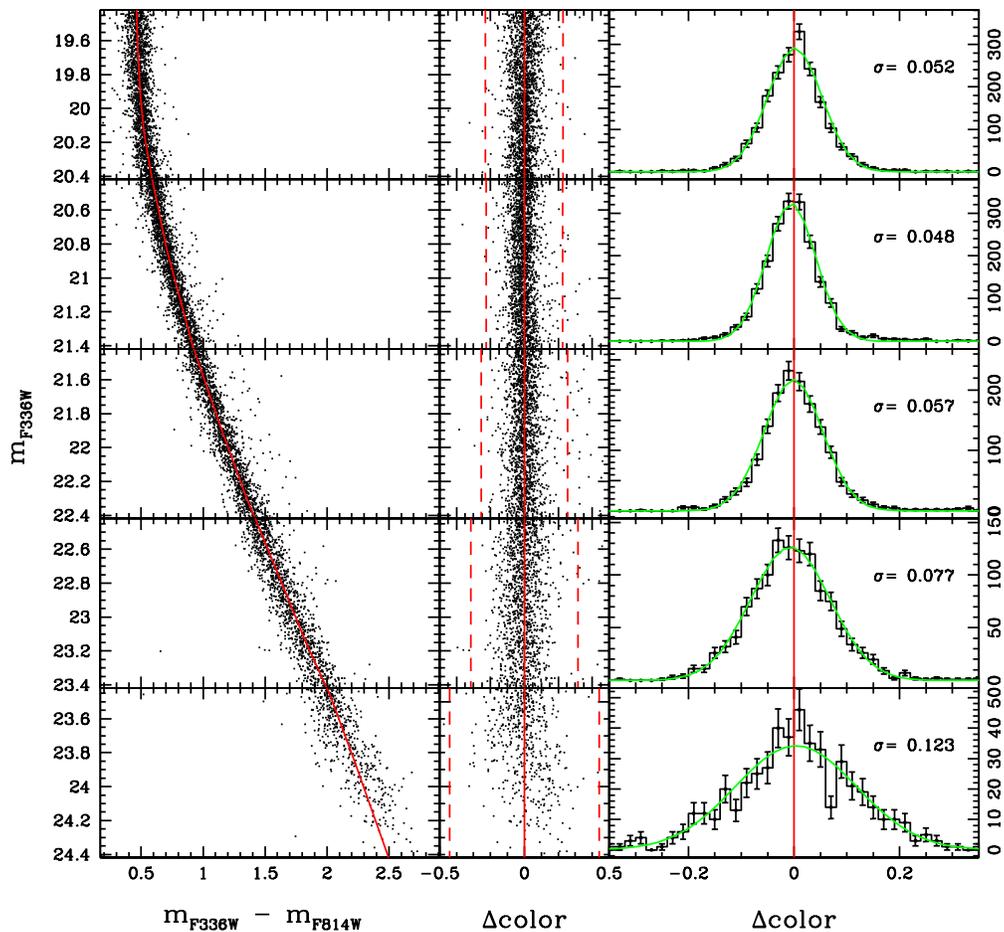}
     \caption{{\it Left}: The same CMD as shown in the right-hand panel
     of Fig.~\ref{f2}, with our MSRL overplotted; {\it Middle}: The CMD,
     rectified by subtraction of the MSRL; {\it Right}: Color
     distribution of the rectified CMD.  The $\sigma$ in the inset are
     those of the best-fitting Gaussians.}
         \label{f4}
   \end{figure}

Fig.~\ref{f4} shows the color distribution of the MS stars.  The left
panel repeats a part of the proper-motion-selected CMD that we showed
in Fig.~\ref{f2}.  In order to derive the main sequence ridge line
(MSRL) we divided the CMD into intervals of 0.2 mag in the F336W band
and computed the median color in each interval.  By fitting these
median points with a spline, we obtained a raw fiducial line. Then for
each star we calculated the difference in color with respect to the
fiducial, and took as the $\sigma$ the location of the 68th percentile
of the absolute values of the color differences. All stars
with a color distance from the MSRL greater than $4\sigma$ were
rejected, and the median points and the $\sigma$ were redetermined.  The
middle panel of Fig.~\ref{f4} shows the same
CMD as the left panel, but after subtracting from each star the MSRL
color appropriate for its F336W magnitude.  Finally, the right panel
shows the histograms of the color distributions in five different
magnitude intervals.

The color distribution is fairly well reproduced by a Gaussian, and
there is no evidence of a split of the MS, such as observed in
$\omega$~Cen or NGC 2808.  Since most of the error in color will come
from the WFPC2 photometry, which will at best have an error of 0.02
mag (Anderson \& King 2000), a 0.01 mag color error in the WFC
(Anderson \& King 2006) will give us an expected color error of 0.025
mags from measuring error alone.  As Fig.~\ref{f4} shows that the MS
broadening can be represented with a Gaussian with $\sigma\sim0.05$,
we obtain an upper limit for the intrinsic color dispersion of the MS
of NGC 1851 of $\sigma_{MS}=0.04$.

%__________________________________________________________________
%__________________________________________________________________
%

%%%
%
\section{Two Stellar populations in NGC 1851}
%
%%%
\label{SGB}

In this section we will derive the basic properties of the two SGBs,
and consider
whether other features in the CMD may indicate the presence
of two stellar populations in NGC 1851.

\subsection{SGB population ratios}

   \begin{figure}
   \epsscale{.80}
   \plotone{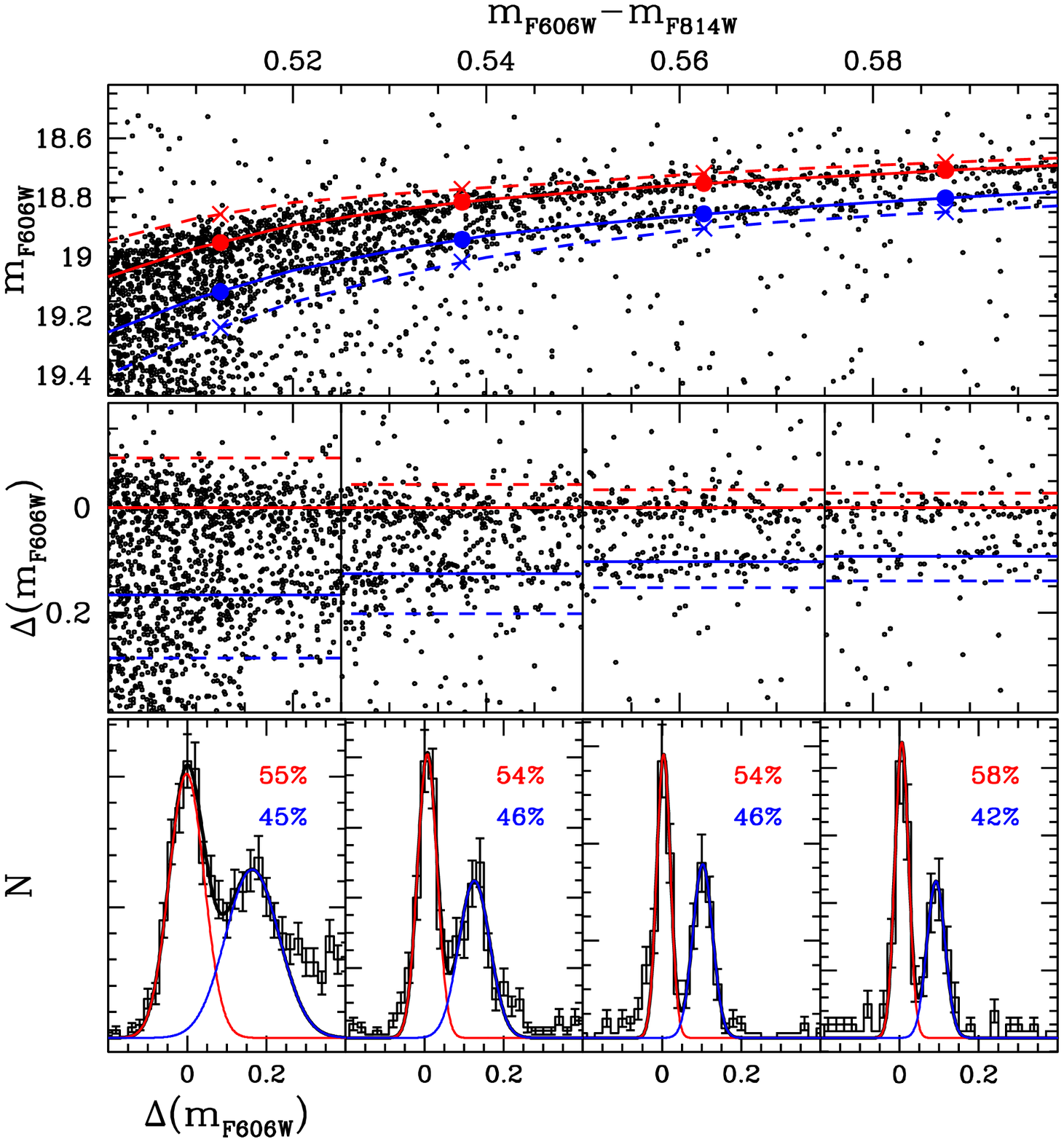}
      \caption{
{\it Top}: F606W vs.\ F606W $-$ F814W diagram zoomed around the SGB. The
red continuous lines show the adopted fiducial line for the bSGB; the
other lines are explained in the text. {\it Middle}: The same diagram,
after subtraction of the bSGB fiducial line.  The lines are explained in
the text.  {\it Bottom}: Distribution in magnitude of the bSGB stars
(red) and fSGB (blue).  The numbers indicate the fraction of stars in
each SGB, as the area of the best-fitting Gaussians, shown in the plot.
(See the text for more details.)
}
         \label{f5}
   \end{figure}

Figure \ref{f5} shows how we fitted Gaussians to the magnitude
distributions of the bSGB and fSGB stars to estimate their number ratio.
The top panel of the figure is a blow-up of the part of the SGB where
the split occurs.  The red line is a fiducial line through the bSGB.  We
drew it by marking the middle magnitude level of bSGB at four equally
spaced points and drawing a line through them by means of a spline fit.
In the middle panel the magnitude of each star has had subtracted from it
the magnitude of this fiducial line at the color of the star.  For the
analysis that was to follow, we divided the color range into four
sections, as shown.

In each color section we estimated the fraction of stars in each of the
SGBs, as follows.  Our aim was to fit the magnitude distributions in the
two branches by a pair of overlapping Gaussians, but we felt the need to
eliminate outliers, especially in the bluest color section.  This we did
by making a preliminary least-squares fit of the Gaussians, using all
of the stars.  We then repeated the solution, but omitting the stars
that lay more than $2\sigma$ above the midpoint of the Gaussian for bSGB
or more than $2\sigma$ below the midpoint of the Gaussian for fSGB
(using in each case the sigma of the corresponding Gaussian).
In the middle panel of Fig.~\ref{f5} the horizontal lines show the
midpoint of the final Gaussians.  The red dashed line runs $2\sigma_b$
on the bright side of the bSGB, and the blue dashed line runs
$2\sigma_f$ on the faint side of the fSGB (where the $\sigma$'s are
those of the best-fitting Gaussians in each color section, fitted to the
distributions of the bSGB and fSGB stars, respectively).  The stars
rejected consist of field stars, objects with poor photometry, or
binaries (brighter than the bSGB).  We have also marked, in the top
panel of Fig. \ref{f5}, the values of the Gaussian centers of bSGB with
large dots and the spline through them with continuous lines.  The
crosses are the Gaussian centers of fSGB lowered by $2\sigma_f$, and the
dashed lines are a spline fitted through the crosses.

The bottom section of Fig.~\ref{f5} shows the histogram of the magnitude
distribution in each color section of the middle panel, and the
best-fitting Gaussians. The numbers reported in the figures give the
percentage of total area under each of the two Gaussians, i.e., the
percentage of bSGB (red) and fSGB (blue) in each interval.

When we added up, for each SGB, the areas under the four Gaussians
fitted to the separate color sections, we found that 55\% of the stars
belong to the bSGB, and 45\% to the fSGB.  Although the expected sigma
of samplings from a binomial distribution of more than 1000 stars, with
a probability of 0.55, is less than 2\%, we estimate that our observed
55/45 split really has an uncertainty that is more like 5\%.  This
estimate is based on the differing results of several different ways of
fitting the Gaussians (e.g., as simple a thing as changing the bin-width
of the histograms).

As for the intrinsic width of each SGB, the dispersions in magnitude of
the two Gaussians in the color sections where the two SGBs are well
separated are of the order of 0.02 mag, consistent with the
uncertainties in our corrections for spatial variation of the PSF.
(Paradoxically, WFC colors are more accurate than WFC magnitudes,
because these corrections are what limits the accuracy of high-S/N
photometry, and in the case of colors the corrections are only for the
difference in behavior of the PSF through two different filters.)

\subsection{Spatial distribution of the SGB stars}

\begin{figure}
   \epsscale{.80}
   \plotone{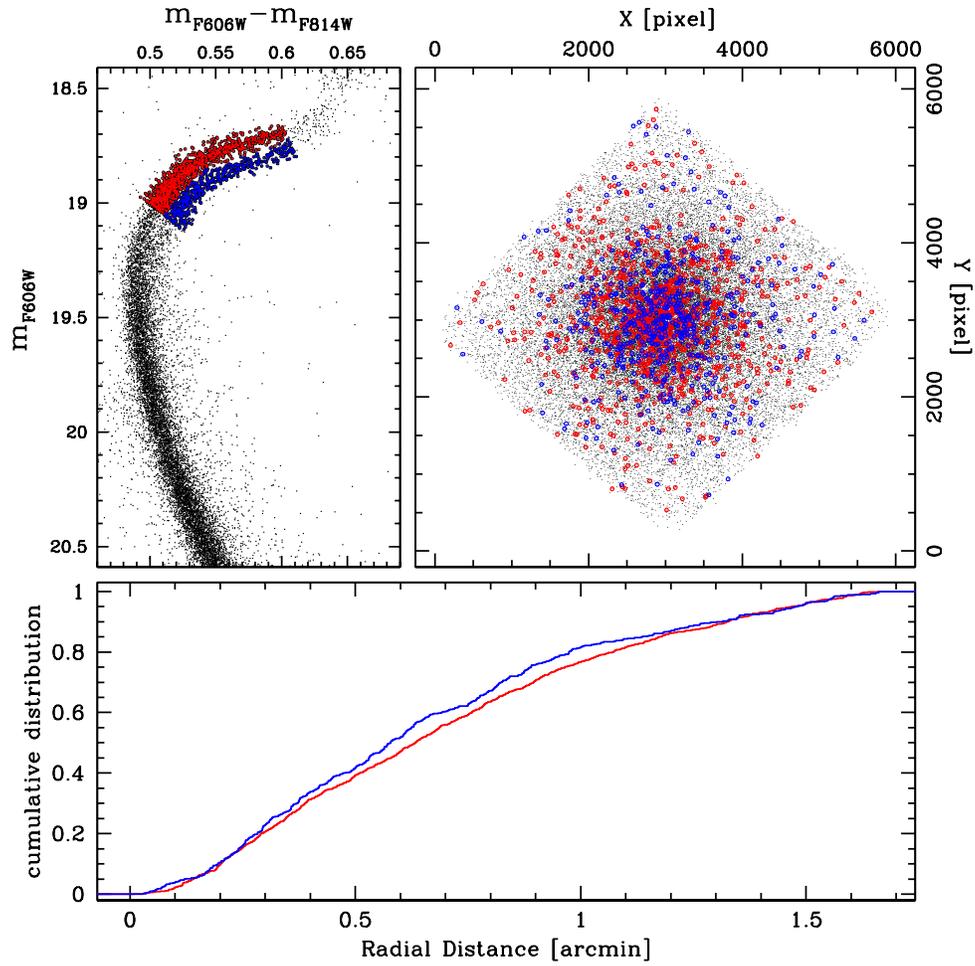}
      \caption{{\it Lower right}: Selection of the bSGB stars (red dots)
      and fSGB stars (blue dots). {\it Lower Left}: Spatial distribution
      of the bSGB and fSGB stars. {\it Top}: Cumulative radial
      distributions of the bSGB and fSGB stars.}
         \label{f6}
   \end{figure}

Figure \ref{f6} shows the spatial distributions of the two SGB
components. 
We select the two SGB subsamples as shown in the
bottom-right panel, and plot the spatial distributions in the
bottom-left panel.  Stars from both groups have similar spatial
distributions.  This is also confirmed by the
cumulative radial distributions shown in the top panel.  The
Kolmogorov-Smirnov statistic shows that in random samplings from the
same distribution a difference this large would occur 11\% of the
time, which is very reasonable for the hypothesis that the two SGBs
have the same distribution.

%__________________________________________________________________
%__________________________________________________________________
%

%%%
%
\subsection{The horizontal branch stars}
%
%%%

\begin{figure}
   \epsscale{.80}
   \plotone{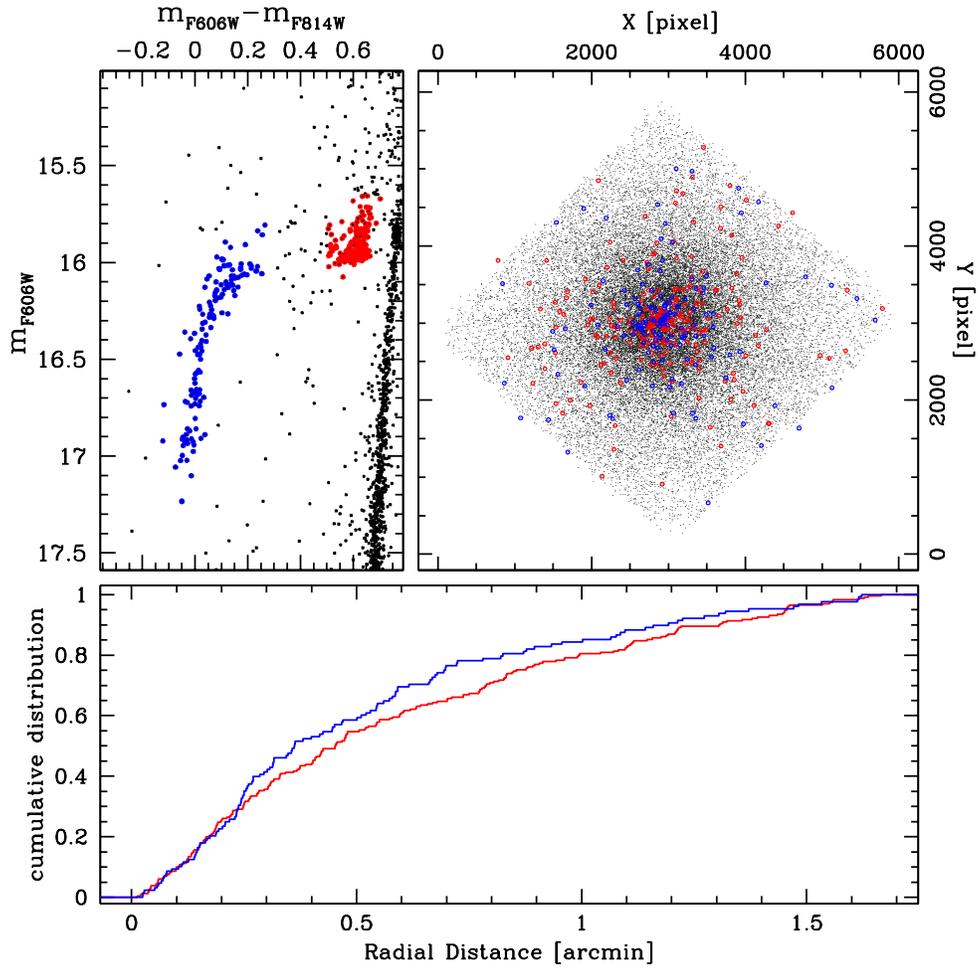}
      \caption{As in Fig.~\ref{f6}, but for the HB.}
         \label{f7}
   \end{figure}

%
% rHB   242 (63%)             younger
% bHB   143 (37%)             older
%
Since the WFC data set included short exposures, we can also examine
the horizontal branch (HB) population.  Figure~\ref{f7} shows the
upper part of the F606W vs.\ F606W $-$ F814W CMD.  The NGC 1851 HB is
clearly bimodal. Interestingly enough, the red HB contains $63\pm7$\%
of the total red + blue HB stars (242 stars), and the blue part
contains $37\pm9$\% of the red + blue HB stars (143 stars).  Note that
some stars populate the RR Lyrae gap. It is difficult to extract RR
Lyrae variables from the small number of exposures that went into
Fig.~\ref{f7}, with each star caught at random phases in each color
band.  In any case, the contribution of the RR Lyraes to the total
population of the HB stars is $<$10\% (see Saviane et al.\ 1998 and
Walker 1992).  As in the case of the bSGB and fSGB, stars from the red
and blue HB groups have similar spatial distributions (see top right
panel of Fig.~\ref{f7}).  This is also confirmed by the cumulative
radial distributions shown in the bottom panel.  The Kolmogorov-Smirnov
statistic shows that in random samplings from the same distribution a
difference this large would occur 17\% of the time, so here too there
is no significant indication of a difference in their radial
distribution. 

Neither for the SGB stars nor for the HB stars is the distribution of
the two samples significantly different.  It would obviously be
desirable to strengthen the statistics by combining the two sets of
counts, but unfortunately this is not possible, because
the incompletenesses in star counts at the HB and at the SGB level are
different.
 
%__________________________________________________________________
%__________________________________________________________________
%

On the basis of the relative numbers of red to blue HB stars and of
bSGB to fSGB populations, 
it is very tempting to connect the bSGB (which includes 55\% of the
SGB stars) with the red HB, and the fSGB (45\% of the SGB stars) with
the blue HB stars.

\section{The ages of the two populations}

The dispersion in the MS in F336W $-$ F814W is less than 0.04 mag.
Such a narrow main sequence implies that either the two SGB
populations have the same metallicities and the same helium abundances
or some fortuitous combination of abundances that results in no MS
split.  Comparison with the theoretical models of Dotter et al.\
(2007) provides an upper limit to a possible dispersion in [Fe/H] of
0.1 dex or in helium abundance of $\Delta Y= 0.026$, if each of the
two is assumed to act alone.  We are not suggesting that such spreads
exist; we merely state upper limits to what might be present.

We note that if we just increase the He abundance, there will be no
appreciable effect on the SGB luminosity. Helium alone cannot explain
the observed SGB split. On the other hand, a population 0.2 dex more
metal rich would have an SGB that is fainter by $\sim0.12$ magnitude,
and therefore it would be able to explain the SGB split. However, such a
high difference in [Fe/H] is ruled out by the narrowness of the MS.  We
have also looked at the color distribution of the red giant branch (RGB)
and found no evidence of a split, and found an RGB color dispersion
consistent 
with the photometric error, confirming that any dispersion in [Fe/H]
alone must be much smaller than 0.2 dex.

Note, however, that increasing [Fe/H] makes the MS redder, while
increasing helium makes the MS bluer.  Thus if, as a mere working
hypothesis, one increases [Fe/H] by 0.2 dex (enough to explain the SGB
split), while changing $Y$ from 0.247 to 0.30, then the two effects
cancel each other out, and the MS would have the same color. The same
would happen on the RGB:\ an [Fe/H] content 0.2 dex higher would move
the RGB to the red, but the RGB would be moved to the original color in
the presence of a He content $Y \sim 0.30$. In this way we would be able
to reconcile the SGB split with the low dispersions in the MS and RGB.
(Clearly, this kind of conspiracy cannot be ruled out till we have
high-resolution spectra to measure the metal content of the NGC 1851
stars.)  However, there is additional observational evidence which seems
to rule out such an unlikely coincidence. A change in [Fe/H] of +0.2 dex
would imply an HB 0.03--0.04 magnitude fainter, while the suggested $Y$
enhancement would end with a bluer HB that is $\sim0.2$ magnitude
brighter.  The resulting sloping of the horizontal branch is not
observed in Fig.~\ref{f7}.

The only remaining possibility of explaining the SGB split is therefore
to assume
that in NGC 1851 there are two stellar populations with similar
metallicity and He content, but with different ages.
Figure~\ref{f5} shows that the two SGBs are separated vertically by
$\sim0.1$ magnitude in F606W. Comparison with the theoretical models of
Dotter et al.\ (2007) shows that this separation corresponds to an age
difference of $\sim 1$ Gyr, with the fSGB 
older than the bSGB.

%__________________________________________________________________
%
\section{Discussion}

In this paper we have shown that the SGB of the Galactic GC NGC 1851 is
split into two distinct branches.
If the split is interpreted in terms of age, its width implies that the
two populations were formed at epochs separated by $\sim1$ Gyr.
As discussed in the previous section, the age difference seems a
possible explanation for the peculiar SGB.

The SGB split is not the only observational evidence of the presence of
more than one population in NGC 1851, though it is probably the clearest
one.  We have already noted that NGC 1851 has a bimodal HB, and that
there might be a connection between the two SGBs and the two HB
sections. The relative frequency of stars in the different branches
implies that the progeny of the bSGB are in the red part of the HB, while
the progeny of the fSGB are on the blue side of the instability strip.
This is what one would expect qualitatively from stellar evolution
models, though an age difference of only 1 Gyr is not enough to move
stars from the red to the blue side of the HB.  Our models indicate
that to move stars from the red to the blue side of the RR Lyrae gap would
require an age difference of $\sim$ 2--3 Gyr.
Some additional parameter, or combination of parameters, must be at work
in order to explain the morphology of the HB of NGC 1851.

There is another relevant observational fact that must be mentioned.
Hesser et al.\ (1982) found that three out of eight of their bright RGB stars
show ``extraordinarily strong'' CN bands. These stars also show
enhanced Sr II and Ba II lines, and lie systematically on the red side
of the RGB.  In other words, $\sim$40\% of their sample of bright RGB
stars contains CNO-processed material.  This material could come from
the interior of these stars through mixing processes, but after what
we have learned from $\omega$~Cen and NGC 2808 we cannot exclude
the possibility that this processed material comes from a
first generation of stars that polluted the gas from which the
CN-strong stars have formed.  This hypothesis would be further
supported if Sr and Ba are also confirmed to be enhanced (as the
spectra of Hesser et al.\ [1982] seem to imply), as these elements
cannot be produced in the low-mass stars presently on the RGB of NGC
1851.

Unfortunately, the results by Hesser et al.\ (1982) are based on a
very limited sample of stars; a more extended spectroscopic
investigation is clearly needed. Meanwhile, it is rather instructive
to look at the CMD published by Grundahl et al.\ (1999), based on
Str\"omgren photometry.  Among the 15 CMDs in their Fig.\ 1, NGC 1851
shows by far the broadest RGB, with some hint of a bimodality.  This
bimodality is not visible in our narrow-color-baseline F606W vs.\
F606W $-$ 814W CMD, but the capability of Str\"omgren photometry to
distinguish stellar populations with different metal content (in
particular CN content) is well known, and the CMD of Grundahl et al.\
(1999) tends to confirm the results of Hesser et al.\ (1982).  We have
investigated the possibility that an enhancement of C or N, or of both
elements, can be the cause of the observed SGB split.  Hesser et al.\
(1982) found that model spectra with [C/A] = +0.2 and [N/A] = +0.5
(where they define [A/H] as the logarithmic relative abundance of all
heavy elements in the theoretical models), or else model spectra with
with [C/A] = 0 and [N/A] = +1.0 could fit the observed spectra equally
well.  A detailed analysis of the effect of C and N overabundances is
beyond the scope of the present paper.  Furthermore, before such an
analysis is justified we will need better abundance constraints, which
would require fitting a larger sample of stellar spectra with modern
atmospheric models.

Here we note that one possible consequence for a second generation of
stars is indeed an increased CNO abundance from mass lost after the
third dredge-up in intermediate-mass AGB stars from an earlier
generation (see discussion in Ventura \& D'Antona 2005 and Karakas et
al.\ 2006).  The spectroscopic results of Hesser et al.\ (1982)
suggest enhanced levels of C and N but do not include an analysis of
the oxygen abundance, due to a lack of measurable features in the
spectra.  (This further underlines the need for a new study.)  It is
however possible that the C + N + O amount is enhanced to some extent,
though at the moment we cannot quantify this enhancement. The CNO
enhancement is an important observational input, as the level of
enhancement would allow us to better identify the possible polluters
from the first generation of stars.

The presence of CNO enhancement would affect the age difference between
the two populations. A CNO enhancement of +0.3 dex would
reduce the age difference between the two SGBs to 300--500 Myr.
Under this hypothesis, we estimate a gap in the SGB of $\sim0.07$ magnitude
due to the composition alone, and an additional 0.03--0.05 mag
displacement due to the age difference, fully accounting for the observed
magnitude difference between the two SGBs.

Larger CNO enhancements would further reduce the age gap between the
two generations of stars, down to a few hundred Myr, as expected in
the intermediate-mass-AGB-star pollution scenario (Ventura et al.\
2002).  We note that if the CNO enhancement in second-generation stars
of NGC 1851 is confirmed, NGC 1851 might differ from other globular
clusters, in which the sum of CNO elements for normal and
self-enriched stars seems to be constant (Cohen \& Melendez 2005).
New spectroscopy of stars in NGC 1851, along with the photometry
presented in this paper and detailed modeling, can significantly
improve our understanding of this intriguing situation.

NGC 1851 is the third globular cluster for which we have direct evidence
of multiple stellar generations.  All three clusters seem to differ in
several important respects, however:

\begin{itemize}

\item
In $\omega$~ Cen, the multiple populations manifest themselves both in
a main-sequence split (interpreted as a bimodal He abundance; see
Bedin et al.\ 2004, Norris 2004, and Piotto et al.\ 2005) and in a
multiplicity of SGBs (interpreted in terms of large age variations,
$\gg 1$ Gyr; see Villanova et al.\ 2007, and references therein),
which implies at least four different stellar groups within the same
cluster.

\item
In NGC 2808, the multiple generation of stars is inferred from the
presence of three MSs (interpreted in this case too in terms of three
groups of stars characterized by different helium contents; see
discussion in Piotto et al.\ 2007), and further confirmed by the
presence of three groups of stars with different oxygen abundances.
It is also consistent with the presence of a multiple HB
(D'Antona \& Caloi 2004).  However, in NGC 2808 the TO-SGB
regions are so narrow that any difference in age between the three
stellar groups must be significantly smaller than 1 Gyr.

\item
In the case of NGC 1851, we have evidence of two stellar groups from
the SGB split, which apparently implies two stellar generations,
formed with a time separation of $\sim 1$ Gyr. This hypothesis is
further confirmed by the presence of a group of RGB stars with strong
CN bands (distinct from the majority of CN-normal RGB stars), and
enhanced Sr and Ba, and possibly the presence of a bimodal HB. In the
case of NGC 1851 there is no evidence of MS splitting, which implies
that any difference in helium or other heavier element content between
the two populations should be small.

\end{itemize}

Apparently, not only are GCs not single-stellar-population objects,
containing stars formed in a single star burst, but the evidence
emerging from the new exciting \hst\ observations presented in this
paper --- as well as from the \hst\ observations of $\omega$ Cen and NGC
2808 --- is that the star-formation history of a globular cluster
can vary strongly from cluster to cluster. We are still far from
understanding what governs the different star-formation histories, and
this is an exciting and challenging question for future investigations.
At the moment, we can only note that the three clusters in which multiple
generations of stars have been clearly identified ($\omega$ Cen, NGC
2808, and NGC 1851), and the two other clusters strongly suspected to
contain more than one stellar generation (NGC 6388 and NGC 6441; see
Caloi \& D'Antona 2007, and Busso et al.\ 2007) are among the ten most
massive clusters in our Galaxy. This evidence suggests that cluster mass
might have a relevant role in the star-formation history of GCs.

We should note finally that the cases of $\omega$ Centauri and M54,
both probably associated with mergers of other galaxies into the Milky
Way, suggest the possibility that all globular clusters that have
complexities in their CMDs are likewise merger remnants.  This is an
attractive speculation, but we think that it does not yet have enough
observational support.

\begin{acknowledgements}
The authors wish to thank the referee, Francesca D'Antona, for useful
discussions.  The USA authors acknowledge the support for Program
number GO-10775 provided by NASA through a grant from the Space
Telescope Science Institute, which is operated by the Association of
Universities for Research in Astronomy, Incorporated, under NASA
contract NAS5-26555.

\end{acknowledgements}

%%%

%__________________________________________________________________
%%%%

\end{document}